\documentclass[aps,prl,twocolumn,superscriptaddress,letterpaper]{revtex4}
\usepackage{amssymb}
\usepackage{amsmath}
\usepackage{graphicx}
\usepackage{natbib}
\UseRawInputEncoding
\usepackage{color}
\usepackage{mathtools}
\usepackage{nicefrac}

\DeclarePairedDelimiter\ket{\lvert}{\rangle}
\DeclarePairedDelimiterX\braket[2]{\langle}{\rangle}{#1 \delimsize\vert #2}
\definecolor{darkblue}{rgb}{0,0,0.5}
\definecolor{lila}{rgb}{0.3,0,0.3}
\definecolor{turq}{rgb}{0,0.1,0.4}
\definecolor{lightblue}{rgb}{0.7,0.7,0.9}
\usepackage{url} % that hyperlinks may be hyphenated correctly
\usepackage[pdftex,
 colorlinks=true,
 backref=page,
 linkcolor=darkblue, % usual links
 filecolor=red,
 citecolor=turq, % for bibliographic
 urlcolor=lila, % for Emails etc
 pdftitle={},
 pdfauthor={},
 pdfsubject={},
 pdfkeywords={},
 pdfpagelabels=true, % damit die Nummerierung stimmt
 breaklinks=false,
 plainpages=false,
 backref=false,
 bookmarks,
 bookmarksnumbered=true]{hyperref}

\begin{document}

\title{Certified Randomness from Remote State Preparation Dimension Witness}
\author{Xing Chen}
\affiliation{3. Institute of Physics, University of Stuttgart and Institute for Quantum Science and Technology (IQ$^{\mbox{ST}}$), Pfaffenwaldring 57, D-70569 Stuttgart, Germany}
\author{Kai Redeker}
\affiliation{Fakult\"at f\"ur Physik, Ludwig-Maximilians-Universit\"at M\"unchen, D-80799 M\"unchen, Germany}
\author{Robert Garthoff}
\affiliation{Fakult\"at f\"ur Physik, Ludwig-Maximilians-Universit\"at M\"unchen, D-80799 M\"unchen, Germany}
\author{Wenjamin Rosenfeld}
\affiliation{Fakult\"at f\"ur Physik, Ludwig-Maximilians-Universit\"at M\"unchen, D-80799 M\"unchen, Germany}
\author{J\"org Wrachtrup}
\affiliation{3. Institute of Physics, University of Stuttgart and Institute for Quantum Science and Technology (IQ$^{\mbox{ST}}$), Pfaffenwaldring 57, D-70569 Stuttgart, Germany}
\affiliation{Max Planck Institute for Solid State Research, Heisenbergstra\ss e 1, D-70569 Stuttgart, Germany}
\author{Ilja Gerhardt}
\affiliation{3. Institute of Physics, University of Stuttgart and Institute for Quantum Science and Technology (IQ$^{\mbox{ST}}$), Pfaffenwaldring 57, D-70569 Stuttgart, Germany}
\email{Corresponding author: i.gerhardt@pi3.uni-stuttgart.de}

%TC:ignore
\begin{abstract}
Randomness in Bell test data can be device-independently certified by Bell's theorem without placing assumptions about the experimental devices. The device-independent randomness has very demanding requirement about the experimental devices and relatively lower output randomness. With the same Bell test data we can extract substantially more randomness without using Bell's theorem. To achieve this goal, we introduce a remote state preparation dimension witness and a semi-device-independent randomness certification model which is based on it. This is one important step towards practical use of Bell test in randomness generation.
\end{abstract}
%TC:endignore
\maketitle

%%A small introduction of random number
Random numbers have a wide variety of applications in daily life~\cite{knuth_1997}. Their use covers gambling, scientific research~\cite{galton_n_1890}, and most importantly, cryptography~\cite{bb84}. Naturally, a given bit string can not be proven to be random~\cite{knuth_1997}, and the \emph{generation process} of a random number is the relevant measure. Quantum-mechanical processes are believed to be the only known source of randomness in nature, thus the generation of a random bit by a quantum mechanical superposition is desirable~\cite{Collantes_rmp_2017}. Many quantum mechanical measurements show a probabilistic outcome~\cite{landau_quantummechanics}, however, this can have other, technical causes~\cite{acin_nature_2016,Chen_sr_2019}, than quantum mechanics.

%%certified random numbers by physics inequality
For a reliable quantum random number, i.e.\ one which directly stems from a quantum process and is free from other noise sources or manipulation, we need to utilize a process which proves its ``quantum nature'' in a measurement. The usage of fundamental physics inequalities can realize this goal. From Bell inequalities~\cite{bell_1964}, for example, the Clauser-Horne-Shimony-Holt (CHSH) inequality~\cite{clauser_prl_1969}, random numbers can be \emph{certified} in a device-independent (DI) way~\cite{pironio_nature_2010,acin_nature_2016,Friedman2018,bierhorst_nature_2018,pan_nature_2018,Knill_2020_pef,Knill_2020_quantum_pef}.

%% other SDI measures.
Although Bell's theorem seems to be the ideal way to certify quantum randomness, this method remains experimentally challenging~\cite{Hensen_nature_2015,giustina_prl_2015,shalm_prl_2015,rosenfeld_prl_2017} and has low randomness output rate~\cite{pironio_nature_2010,bierhorst_nature_2018}. Therefore, other \emph{semi-device-independent} randomness certification methods have been developed. \emph{Semi-device-independent} (SDI) means that the raw measurement outcomes reveal their quantumness, if a certain number of assumptions of the experimental setup can be guaranteed~\cite{assump}. Among these we find the Kochen-Specker inequality~\cite{ks_1967,um_sr_2013_certified_random_kcbs} and the dimension witness~\cite{bowles_prl_2014,Jerger_nature_2016,lunghi_prl_2015}.

\begin{figure}[!htb]
\center{\includegraphics[width=0.5\textwidth]{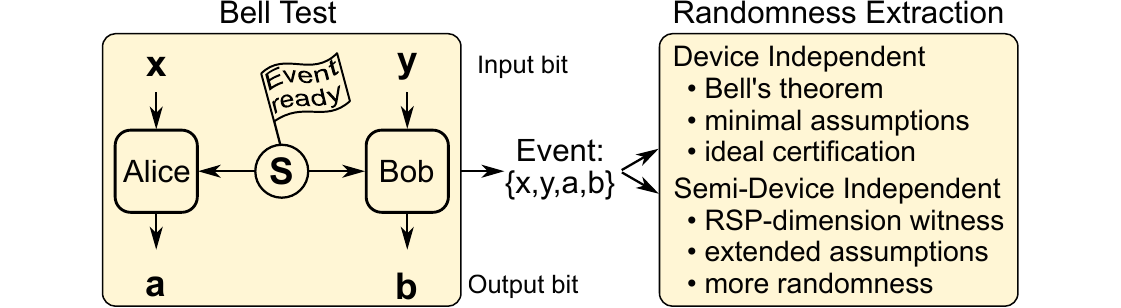}}
\caption{A Bell test involves two physically separated systems, and two given input bits $x, y$ generates outcomes $a, b$. The Bell correlation value $S$, allows in a DI scenario to quantify the amount of randomness; another scenario is to extract SDI randomness, when remote state preparation (RSP) dimension witness is utilized.}
\label{fig:entropy_extraction}
\end{figure}

%%From DI random number to SDI random number
Here, we certify and extract the randomness generated in a loophole-free Bell test~\cite{rosenfeld_prl_2017}, in a DI and a SDI manner~\ref{fig:entropy_extraction}. The randomness is bounded by min-entropy\cite{min_entropy_def}. For the DI approach we use the prior analysis from~\cite{pironio_nature_2010} to quantify the entropy in the data. For the SDI approach, we introduce a remote state preparation dimension witness. This allows for a significantly higher output rate of random bits.

\emph{CHSH scenario}---For the CHSH inequality~\cite{clauser_prl_1969}, an experiment with pairs of particles and two parties, Alice and Bob, is considered. In each round of the experiment, each party receives one particle of a pair and performs a local measurement on it, using one out of two measurement settings. The choice of the local measurement settings depends on the randomly chosen binary input $x$ for Alice and $y$ for Bob. The measurements produce a binary output $a$ for Alice and $b$ for Bob (Fig.~\eqref{fig:entropy_extraction}).

%%introduce CHSH inequality
The correlation value $S$ of the CHSH inequality is $S=|\sum_{x,y}(-1)^{xy}\left[P(a=b|xy)-P(a\neq b|xy)\right]|$~\cite{chsh_value}, where $|.|$ denotes the absolute value, and $P(a=b|xy))$ (or $P(a\neq b|xy)$) is the probability that output $a=b$ (or $a\neq b$) when the measurement settings $(x,y)$ are chosen. For all local realistic theories, $S$ cannot exceed the maximum value of 2. In contrast, quantum mechanics allows the value of $S$ to be between $2$ and $2\sqrt{2}$~\cite{bell_1964,chsh_prl_1969}.

%% Experimental part, remote Rb atoms
The aforementioned experiment is performed as a loophole-free Bell test~\cite{rosenfeld_prl_2017}. In this Bell test, Alice and Bob each operate an atom trap for a single rubidium atom. The traps, separated by 398~m, are independently operated, comprising an own laser system and control electronics. The atomic qubits are encoded in the $m_{F}=\pm1$ Zeeman sub-level of the $5S_{\nicefrac{1}{2}},F=1$ ground state, with $\left|\uparrow\right\rangle _{z}$ corresponding to $m_{F}=+1$, $\left|\downarrow\right\rangle _{z}$ corresponding to $m_{F}=-1$.% To generate entanglement between Alice and Bob's remote atoms, an atom-photon entanglement and a subsequent Bell state measurement (BSM) on the two emitted photons was used~\citep{Zukowski1993,Pan1998}.

%% Atom Photon entanglement explained
For the creation of the entangled atom-photon pairs, each atom is excited to the $5P_{\nicefrac{3}{2}},F'=0,m_{F}=0$ state via a short laser pulse. The subsequent spontaneous emission yields to a photon whose polarization is entangled with the atomic qubit state. Both photons are coupled into single mode fibers and are guided to a BSM setup, where two photon interference on a fiber beam splitter together with photon polarization analysis is employed to project the photons on two out of the four possible Bell states. The photonic measurement heralds the creation of one of the entangled atom state $\left|\Psi^{\pm}\right\rangle =1/\sqrt{2}\left(\left|\uparrow\right\rangle_{x}\left|\downarrow\right\rangle _{x}\pm\left|\downarrow\right\rangle _{x}\left|\uparrow\right\rangle _{x}\right)$, where $\left|\uparrow\right\rangle_{x}=(1/ \sqrt{2})\left(\left|\uparrow\right\rangle_{z}+\left|\downarrow\right\rangle_{z}\right)$ and $\left|\downarrow\right\rangle_{x}=(i / \sqrt{2})\left(\left|\uparrow\right\rangle_{z}-\left|\downarrow\right\rangle_{z}\right)$.

%% explanation of an event, experimental run, x, y, a and b. How many events? How big Bell violation?
After entanglement is created between Alice and Bob, they start a fast atomic state measurement process based on state selective ionization and subsequent detection of the ionization fragments. The measurement setting is determined by the polarization of a laser pulse exciting the atom before ionization. For the choice of the setting each party employs a QRNGs outputting freshly generated random bits on demand. The total time needed from the generation of the input  $x$ or $y$ to receiving the output $a$ or $b$ is less than 1.1~$\mu s$, together with a separation of the atom traps of 398~m this enables for space-like separation of the measurements~\cite{Larsson_2014}. Thus, the experiment enforced the assumptions made for deriving the CHSH inequality. In total, 55568 rounds were recorded, 27885 with the $\ket{\Psi^+}$ prepared and 27683 with the $\ket{\Psi^-}$.

%% The DI meaning
\emph{DI certification protocol}---As outlined above, device-independence can be linked to the violation of Bell inequality~\cite{colbeck2009quantum}. This implies, as long as Bell inequality is guaranteed to be violated, true randomness can be generated.

%%the assumption of DI randomness
Although the randomness certified by Bell's theorem \emph{can} be device independent, we still need some extra assumptions to bound the randomness in this model~\cite{pironio_nature_2010}: (1) the remote parties perform local and independent measurements on their ideally space-like separated (=perfectly shielded) devices; (2) the measurement settings (x,y) are not determined beforehand and are unpredictably chosen; (3) the measurement process is described by quantum mechanics, and nature is not e.g.\ pre-determined as a whole. In a loophole-free Bell test the assumption (1), which is required for a loop-hole free Bell test is fulfilled. Assumption (2) means that the $i-th$ input $x_i$ and $y_i$ are not known to the experimental devices until the $i-th$ run of the experiment.

In~\cite{pironio_nature_2010}, the marginal guessing probability $p(a|x)$ had been connected to the correlation value $S$ of the CHSH inequality. This allows to bound the entropy of the output data to 1 bit per event when $S=2\sqrt{2}$. Due to the finite data size, the confidence level is introduced. Using the analytical model in~\cite{pironio_nature_2010}, and taking the confidence level as 0.99~\cite{pironio_nature_2010,lunghi_prl_2015}, the min-entropy in our Bell test data can be quantified. For the $\ket{\Psi^+}$ state, the data resulted in $S=2.085$, and with a total number of events $n=27885$, no DI randomness can be certificated for this entangled state. Performing the same task for the 27683 events from the $\ket{\Psi^-}$ state, the value of $S$ amounts to $2.177$. The min-entropy of the DI randomness amounts to 531 bits, which is much smaller than our SDI randomness as we show in the following text.

%%Bell test %DI model, no efficient use of data, SDI
\emph{SDI certification protocol}---For the certification of the randomness from Bell test data, Bell inequalities must be violated. Unfortunately, the loophole free Bell experiments~\cite{Hensen_nature_2015,shalm_prl_2015,giustina_prl_2015,rosenfeld_prl_2017}, did not reach the maximally allowed values for quantum mechanics. Thus only a small amount of randomness per round can be certified in the DI manner. As we have shown above, sometimes no randomness can be certified~\cite{lowbell}. However, when we introduce additional assumptions and leave the DI scenario, this situation changes. In order to build the experimental devices, it is necessary to have some knowledge about the way they function e.g.\ the devices are error prone but not maliciously built. This knowledge allows for a higher bound of the randomness per event for the same experiment. Using a \emph{dimension witness} is one possible way for such a higher bound of the randomness.

%%dimension witness
The idea of dimension witness was first introduced in~\cite{brunner_prl_2008}. After this pioneering paper, a substantial number of studies have been performed on this concept~\cite{gallego_prl_2010,li_pra_2011,pawlowski_prl_2011,li_pra_2013,bowles_prl_2014}.

%%prove our system is a 2-dimensional quantum representation
Before applying the dimension witness to the Bell experiment, we first show that the experiment admits a $2$-dimensional quantum representation~\cite{brunner_prl_2008,gallego_prl_2010,bowles_prl_2014}.

%%define x'
In our Bell test, there are two inputs $x,y$ and two outputs $a,b$. When Alice does one of her two measurements, the entangled state of Bob's side will randomly collapse into one specific state. Since Alice has two different measurement settings and each one has two different measurement results, when she does her two measurements randomly multiple times, Bob's side will get four quantum states. These four quantum states in Bob's side are represented as $x'$, and $p(b|x',y)=p(b|x,a,y)=p(ab|xy)/p(a|x,y)$. Since (details in supplementary)
\begin{equation}
\label{eqn:pabxy_to_dw}
p(b|x',y)=\mathrm{Tr}\left(\rho_{a|x}M^B_{b|y}\right)
\end{equation}
%%connect x' with (x,a)
where $\rho_{a|x}$ is the state on Bob's side when Alice performs her measurement $x$ and gets a result $a$. $M^B_{b|y}$ is the measurement operators in Bob's side. $\rho_{a|x}$ and $M^B_{b|y}$ are acting on $\mathbb{C}^{2}$ (a 2-dimensional complex coordinate space). So, $p(b|x',y)$ admits a $2$-dimensional quantum representation~\cite{2dimmore}. This shows that we can use the two-dimensional dimension witness~\cite{bowles_prl_2014} to quantify the quantumness in our Bell test.

%%find the right dimension witness
Different kinds of dimension witnesses can be used in a $2$-dimensional quantum representation. The dimension witness we used here is was introduced in~\cite{bowles_prl_2014}. The advantage of this nonlinear dimension witness is that it can be used for non-convex set and is robust to technical imperfections. Most importantly, it can be used to certify randomness~\cite{bowles_prl_2014}. It is defined as
\begin{equation}
\label{eqn:dw_definition}
W=
\begin{vmatrix}
p(1|0,0)-p(1|1,0)&p(1|2,0)-p(1|3,0)\\
p(1|0,1)-p(1|1,1)&p(1|2,1)-p(1|3,1)
\end{vmatrix}\, ,
\end{equation}
%% ??
where $p(b|x',y)$ is defined in Eqn.~\eqref{eqn:pabxy_to_dw}, and the result $b$ is chosen as ``1'' in the above definition. The definition equation of the dimension witness here is the same as in~\cite{bowles_prl_2014}, but the state $x'$ differs. Here, the state $x'$ is in Bob's side, but its preparation is completed by the projective measurement of Alice, so the state $x'$ is remotely prepared~\cite{bennett_prl_2001,bennett_ieee_2005}. In order to emphasize this difference, we name it as remote state preparation (RSP) dimension witness.

%% what is done in the next paragraph....
From the above definition, the RSP-dimension witness $W_B$ for Bob's side is constructed. Similarly, $W_A$ can be constructed for Alice's side. We define $W_{\mathrm{rsp}}=\min\{W_A,W_B\}$, and use $W_{\mathrm{rsp}}$ as the RSP-dimension witness in the following model. The RSP-dimension witness captures the quantumness of the preparation and measurements in our Bell test. If the preparations are classical, one has $W_{\mathrm{rsp}}=0$, while a quantum preparation and measurement leads to $0< W_{\mathrm{rsp}}\leq 1$.

%%relationship between $S$ and $W$
Although, $S$ and $W_{\mathrm{rsp}}$ are based on the same experimental data, they are not directly linked: $S$ cannot be used to calculate the value of $W_{\mathrm{rsp}}$, it only affects the lower limit of $W_{\mathrm{rsp}}$. For example, when $S=2$, $W_{\mathrm{rsp}}\in\left[0,1\right]$, and when $S=2\sqrt{2}$, $W_{\mathrm{rsp}}=1$.

%%assumptions for SDI conditions
Before using the RSP-dimension witness to bound the randomness generated during the experiment, we discuss the required assumptions. As before, we require, that the above (DI-) assumptions (1, 2, 3) hold true. Besides, there are some extra assumptions~\cite{lunghi_prl_2015}: (4) the information in the measurement results of each side is contained in a two-dimensional quantum subspace; (5) the system is memoryless and subsequent outcomes are not directly correlated.

%%explain assumption (1) in sdi model, and no need of post selection
In the RSP-dimension witness model, it is important that the state preparation and measurement are independent from each other. And this requirement is fulfilled by assumption (1) and (2) in the following way: Let us take $W_B$ as an example. For $W_B$, the states $x'$ in Bob's side are remotely prepared by Alice's quantum measurements, which cannot be affected by Bob's device or measurement, so the states $x'$ are independent from Bob's device and measurement $y$. This independence requirement is naturally fulfilled by our loophole-free Bell test~\cite{rosenfeld_prl_2017}. The prepared states $x'$ might be affected by Alice's device, but that is not a problem for our model. Since if $x'$ is affected by the device in Alice's side, the state in Bob's side will not be properly prepared, and the value of RSP-dimension witness will be decreased. Thus, the independence of $x'$ is quantified by the value of RSP-dimension witness, no post-selection is needed to increase the independence of $x'$. Assumption (1) also implies that the experimental devices do not have any pre-established correlations among each other; this also indicates that the devices that are used to generate the input strings $x,y$ are not correlated with the measurement devices. Subsequently, $x,y$ can be pseudo-random numbers, as long as they are independent from each other and the measurement apparatus.

%%assumption more explanation
Assumption (4) means that the information contains in the measurement result of measuring $x'$ does not exceed 1~bit, a possible violation would be that the information about $x'$ is duplicated by or correlated with extra qubits. The assumption (4) can be relaxed by the space-like separation of Alice and Bob in our Bell test. In general, a bipartite entangled state shared between Alice and Bob has two different measurement results in each side with one measurement setting ---it can be described by a qubit. This does not mean the entangled state shared between Alice and Bob has to be confined in a two-dimensional Hilbert space, it only means with the measurement does by Alice or Bob to it, the results information in contained in a qubit.

%%%
As for the given experiment, the state preparation of the RSP-dimension witness is independently completed by two sides: one side performs the measurement and the other side gets the state simultaneously. Under space-like separation, when the state is prepared by one side, the measurement is performed outside the light cone of state preparation. Thus, it is impossible for the state preparation devices to send extra qubits of the prepared states to the measurement devices without lowering the values of $W_{\mathrm{rsp}}$~\cite{clnoise}. As long as $W_{\mathrm{rsp}}>0$, the remote measurements are exceeding a classical correlation.

%%get the guessing probability under SDI conditions
Since the inputs $x$ and $y$ are independent from each other, and in the experiment, different choices of the measurement settings are uniformly random, thus each combination of $x$ and $y$ occurs with probability $1/4$~\cite{lunghi_prl_2015}. Then, the guessing probability $p_{\rm{guess}}$ of $p(ab|xy)$ is (more intermediate steps are shown in supplementary)
\begin{equation}
\label{eqn:pgabxy}
\begin{split}
&p_{\rm{guess}}(ab|xy)\\
&=\frac14\sum_{x,y}\max_{a,b} p(ab|xy)\\
&\leq\max_{x,a}p(a|x)\frac12\sum_y\max_{x,a,b} p(b|(x,a),y)\\
&\leq\left(\frac{1+\sqrt{1-W^2_{\mathrm{rsp}}}}{2}\right)\frac12\left(1+\sqrt{\frac{1+\sqrt{1-W_{\mathrm{rsp}}^2}}{2}}\right)\, .
\end{split}
\end{equation}
%%conditional min-entropy
As we can see, the equation of guessing probability $p_{\rm{guess}}(ab|xy)$ from $W_{\mathrm{rsp}}$ is not the same as the one from~\cite{lunghi_prl_2015}. The difference is caused by $\max_{x,a}p(a|x)$, which represents the quantum measurement from the state preparation process.

\begin{figure}[!htb]
\center{\includegraphics[width=0.5\textwidth]
  {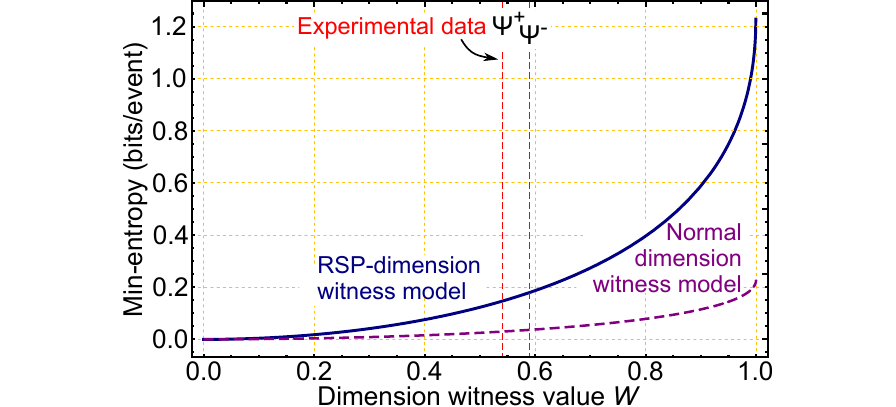}}
\caption{Output randomness utilizing the dimension witness. The nonzero RSP-dimension witness $W_{\mathrm{rsp}}$ gives us a new perspective to bound the randomness in the experimental data. The blue curve displays the randomness certified by the $W_{\mathrm{rsp}}$, while the dashed purple curve represents the randomness certified by the previously defined dimension witness~\cite{bowles_prl_2014}. Clearly, the combination of remote state preparation and the dimension witness increase the bound of randomness per event, as compared to a normal dimension witness certification model in~\cite{lunghi_prl_2015}.}
\label{fig:certified_randomness_dimension_witness}
\end{figure}

%%conditional min-entropy
The conditional min-entropy $H_{\infty}(AB|XY)$ in this situation is $H_{\infty}(AB|XY)=-\mathrm{log}_2 p_{\rm{guess}}(ab|xy)$. This equation allows us to bound the randomness in the experimental data in a SDI way. The randomness per event from the RSP-dimension witness model is depicted in Fig.~\eqref{fig:certified_randomness_dimension_witness}. Compared to~\cite{lunghi_prl_2015}, the introduction of quantum measurements in the state preparation process gives us a significant advantage to bound the randomness in our experimental data. For instance, the maximum certified randomness in our model is 1.23 bits per event, which is significantly larger than the previous dimension witness model~\cite{lunghi_prl_2015}.

\noindent
The presented RSP-dimension witness model can certify substantially more randomness in the Bell test from a different perspective. Only a few more extra assumptions are required for this. Moreover, when $S$ is below the classicality bound 2, the $W_{\mathrm{rsp}}$ can still be larger than 0, see example below.

%%small advantage from our dimension witness
In a practical Bell test, because of the imperfect measurements or entangled states, the Bell inequality might not be violated. In this case, no randomness can be certified by previous models~\cite{pironio_nature_2010,acin_prl_2012_randomness_nonlocality,li_pra_2012,bierhorst_nature_2018,Friedman2018}. With the RSP-dimension witness model, randomness in the experimental data can be certified without using Bell's theorem. For instance, let Alice and Bob share a Bell state, then they measure it with two identical measurement settings $\widehat{x}$ and $\widehat{z}$ at each side (which corresponds to the BBM92 quantum key distribution scheme ~\cite{bennett_prl_1992}). In this case, Bell inequalities will not be violated, but the bound of randomness is 1.23 bits per event data from $W_{\mathrm{rsp}}$.

This example manifests that by rotating Alice and Bob's measurement basis, the RSP-dimension witness is not changed, and it also shows the robustness of our SDI protocol. The corresponding $S$ value would of course be decreased. This demonstrates that, without using Bell's theorem, randomness in the Bell test data can still be certified.

We further consider the following 2-qubit Werner state~\eqref{eqn:wernerstate} as an example,
\begin{equation}
\label{eqn:wernerstate}
\rho_z=z|\Psi^+\rangle\langle\Psi^+|+\frac{1-z}{4}\textbf{I}\, ,
\end{equation}
where $0\leq z\leq 1$, is the noise parameter. For this state the relationship between $W_{\mathrm{rsp}}$ and $S$ can be derived. On one hand, the relationship between $z$ and $S$ is $S=2\sqrt{2}z$. On the other hand, following~\cite{bowles_prl_2014}, the relationship between $z$ and the RSP-dimension witness is derived as $W_{\mathrm{rsp}}=z^2$. Subsequently, the relationship between $W_{\mathrm{rsp}}$ and $S$ is calculated as $W_{\mathrm{rsp}}=S^2/8$. From this relationship we can also see that $W_{\mathrm{rsp}}$ is nonzero when $0<S\leq2$. This shows again that, without using Bell's theorem, randomness in the Bell test data can be certified.

Our SDI model and the DI model in~\cite{pironio_nature_2010} both utilize quantum correlations to certify the quantum randomness. The DI model uses the correlation between the measurement results to form a CHSH inequality, the violation of CHSH inequality guarantees that the randomness is certified. In the SDI model, the correlation between Alice and Bob's measurement results is not quantified by Bell's theorem, but by our nonlinear RSP-dimension witness. This RSP-dimension witness can quantify the quantum correlation which cannot be quantified by the CHSH inequality, such as the case in BBM92 scenario and 2-qubit Werner state.

%%Apply model to experimental data
Next we apply SDI model to bound the randomness produced in the Bell test~\cite{rosenfeld_prl_2017} and then extract the randomness with hashing functions. In the following randomness extraction, due to the finite data size, the confidence level of the model and the error of hashing functions~\cite{hashing_error} are introduced. The confidence level is again taken as 99\%, and the hashing error is chosen as 0.001. We use universal hashing functions to extract the bounded randomness, see supplementary for details. Considering the $\ket{\Psi^+}$ state, the collected data resulted in $S=2.085$, and with a total number of events $n=27885$. We calculate the RSP-dimension witness value for this entangled state as $W_{\mathrm{rsp}}=0.542$. The SDI randomness extracted in all events amounts to 3821~bits, which is a tremendous improvement compare to the 0 bits in DI model of~\cite{pironio_nature_2010}.

Performing the same task for the 27683 events from the $\ket{\Psi^-}$ state, the value of $S$ amounts to $2.177$, and the RSP-dimension witness value is $W_{\mathrm{rsp}}=0.591$. The extracted SDI randomness amounts to 4660 bits, which is much larger than 531 bits DI randomness~\cite{value_explain}.

\emph{Conclusion}--We have presented two methods to bound the randomness in our Bell test data. The DI model from~\cite{pironio_nature_2010} is based on Bell's theorem, and its applicability holds especially for the CHSH-variant of the test~\cite{chsh_prl_1969}. For all the 55568 events data, the min-entropy of the DI randomness is 531 bits, which is $10^{-2}$ bits per event data. An extended RSP-dimension witness model is newly designed in this work for the same version of Bell test. For all 55568 events data, the total min-entropy of the extracted SDI randomness is 8481 bits, corresponding to 0.153 bits per event data, which is significantly higher than $10^{-2}$ bits per event data.

Our RSP-dimension witness model improves the bound of the randomness from the data tremendously without using Bell's theorem and it is still possible to extract randomness with this model when the Bell inequality is not violated as shown in the paper. Of course, the model offers weaker security guarantees for randomness than the DI model, but it is still certified randomness under SDI conditions. Also, the requirements in the SDI model can be fulfilled by standard technologies, which are much less complex than the loophole-free Bell test.


\begin{thebibliography}{46}
\bibitem[{ass()}]{assump}
\bibinfo{note}{For the semi-device-independent protocols mentioned in this
  letter, the assumptions about the experimental devices should be general,
  which means they are not supposed to characterize the devices in details, and
  they do not belong to source-independent or measurement-device-independent
  protocols.}
\bibitem[{chs()}]{chsh_value}
\bibinfo{note}{In order to allow for a violation of the CHSH inequality, for
  certain states and measurement settings combinations, a permutation of the
  values of $a$ and $b$ might be necessary,}.
\bibitem[{low()}]{lowbell}
\bibinfo{note}{This maybe caused by (i) the finite amount of events or (ii)
  simply that Bell inequality is not violated}.
\bibitem[{2di()}]{2dimmore}
\bibinfo{note}{The entangled state shared between Alice and Bob can be in
  higher dimension, it does not need to be reduced to a lower dimension. Alice
  and Bob only measure this entangled state in a 2-dimensional Hilbert space.
  Thus, the input and output data will form a distribution $p(b|x',y)$, which
  admits a $2$-dimensional quantum representation. As long as the input $x,y$
  and output $a,b$ form a 2-dimensional quantum representation, regardless of
  its non-convexity and its origins, the nonlinear dimension witness can be
  applied.}
\bibitem[{cln()}]{clnoise}
\bibinfo{note}{This means, the extra qubits sent out by the state preparation
  devices can be treated as classical noise in this situation.}
\bibitem[{val()}]{value_explain}
\bibinfo{note}{The 531 bits randomness is the bounded randomness from the DI
  model, not considering the hashing error. If we take 0.001 hashing error into
  consideration, this value will decrease to 495 bits.}
\end{thebibliography}
\end{document}